\documentclass[%
 reprint,
 amsmath,amssymb,
 aps,
floatfix,
superscriptaddress]{revtex4-2}

\usepackage{graphicx}
\usepackage{dcolumn}
\usepackage{bm}
\usepackage[colorlinks,linkcolor=blue,anchorcolor=blue,urlcolor=blue, citecolor=blue]{hyperref}
\usepackage{float}
\usepackage{physics}
\usepackage{xcolor}
\usepackage{lipsum}  
\usepackage{mathtools}
\usepackage{subfig}

\usepackage{dsfont}

\usepackage{etoolbox}

\newcommand{\pt}[1]{\left( #1 \right)}
\newcommand{\pq}[1]{\left[ #1 \right]}
\newcommand{\pg}[1]{\left\{ #1 \right\}}

\newcommand{\nuc}{${}^{13}$C }
\newcommand{\cnot}{$\text{C}_n\text{NOT}_e$ }

\begin{document}

\title{Gate Optimization via Efficient Two-Qubit Benchmarking for NV Centers in Diamond}

\author{Alessandro Marcomini}
\email{amarcomini@vqcc.uvigo.es}
\affiliation{Peter Grünberg Institute -- Quantum Control (PGI-8), Forschungszentrum J\"ulich GmbH, D-52425 Germany}
\affiliation{Dipartimento di Fisica e Astronomia ``G. Galilei", Università di Padova, I-35131 Padova, Italy}
\affiliation{Vigo Quantum Communication Center, University of Vigo, Vigo E-36310, Spain}
\affiliation{Escuela de Ingeniería de Telecomunicación, Department of Signal Theory and Communications, University of Vigo, Vigo E-36310, Spain
}
\affiliation{AtlanTTic Research Center, University of Vigo, E-36310, Spain
}%
\author{Philipp J. Vetter}
\affiliation{Institute for Quantum Optics, Ulm University, Albert-Einstein-Allee 11, 89081 Ulm, Germany}
\affiliation{Center for Integrated Quantum Science and Technology (IQST), Ulm University, 89081 Ulm, Germany}
\author{Tommaso Calarco}
\affiliation{Peter Grünberg Institute -- Quantum Control (PGI-8), Forschungszentrum J\"ulich GmbH, D-52425 Germany}
\affiliation{Institute for Theoretical Physics, University of Cologne, D-50937 Germany}
\affiliation{Dipartimento di Fisica e Astronomia, Università di Bologna, 40127 Bologna, Italy}
\author{Felix Motzoi}
\affiliation{Peter Grünberg Institute -- Quantum Control (PGI-8), Forschungszentrum J\"ulich GmbH, D-52425 Germany}
\affiliation{Institute for Theoretical Physics, University of Cologne, D-50937 Germany}
\author{Fedor Jelezko}
\affiliation{Institute for Quantum Optics, Ulm University, Albert-Einstein-Allee 11, 89081 Ulm, Germany}
\affiliation{Center for Integrated Quantum Science and Technology (IQST), Ulm University, 89081 Ulm, Germany}
\author{Matthias M. Müller}
\affiliation{Peter Grünberg Institute -- Quantum Control (PGI-8), Forschungszentrum J\"ulich GmbH, D-52425 Germany}

\begin{abstract}
    High-fidelity gate implementation requires sophisticated control pulses that steer the quantum system to undergo the desired transformation. 
    Quantum Optimal Control allows to derive these control pulses in an open-loop fashion based on numerical simulations.
    However, their precision can be limited by incomplete knowledge of the system. 
    Closed-loop optimization overcomes this limitation by incorporating feedback from measurements, provided a suitable and efficient measure of the gate performance can be defined.
    In this article, we present an efficient method to evaluate the performance of a two-qubit gate by preparation and measurement of only two quantum states, enabling experimental closed-loop optimization with a metric previously believed to be limited to open-loop control.
    We tailor the approach to nitrogen-vacancy centers in diamond and, through numerical simulations, demonstrate how the method can optimize a two-qubit gate while reducing the number of required measurements by two orders of magnitude compared to standard process tomography under realistic experimental settings.

\end{abstract}

\maketitle


\newpage

\section{Introduction}

Engineering fragile quantum systems toward interesting scientific phenomena and technological applications requires precise control over the system dynamics. Quantum Optimal Control (QOC) has been demonstrated to enable this by engineering time-dependent control pulses that steer the quantum state to undergo a desired transformation~\cite{Brif2010,Glaser2015,Koch2022,Khaneja2005,Theis2018,Mueller2022,Rembold2020}.
A faithful numerical description of the system dynamics can be used to calculate control pulses in an open-loop fashion. However, the precision of the controlled operation is intrinsically limited by the precision of the model. To overcome this limitation, feedback from the experiment can be used to evaluate and improve the performance of the control pulses in a closed-loop fashion~\cite{Rosi2013,Frank2017,Heck2018,preti2022continuous,werninghaus2021leakage,gao2025ultrafast,Oshnik2022,Mueller2022,Rossignolo2023,Vetter2024}. A versatile control algorithm that allows to swiftly switch between open- and closed-loop mode builds on the expansion of the control pulse in a suitable basis, the so-called (dressed) Chopped RAndom Basis (dCRAB) algorithm~\cite{Doria2011,Caneva2011,Rach2015,Mueller2022}. A ready-to-use version of the dCRAB algorithm is available in the Quantum Optimal Control Suite (QuOCS) software~\cite{Rossignolo2023}. Since the evaluation of the control-pulse performance can be time-consuming, especially via measurements in the closed-loop mode, it can be advantageous to exploit symmetries in the quantum system or the target operation to reduce the effective degrees of freedom that have to be controlled and measured, e.g. via the identification of invariants~\cite{Mueller2011,Watts2015,Goerz2015,Goerz2017,Baran2026} or of specific states that are most suitable for benchmarking~\cite{Goerz2014,Mueller2014,Reich2013,Goerz2021,Petruhanov2023,Romer2025}.

In this article we combine the approach of closed-loop control with the reduced evaluation protocol for controlled quantum gates first outlined in Ref.~\cite{Goerz2014}, and extending the work carried out in Ref.~\cite{MarcominiMScThesis}. The protocol we derive allows to optimize a two-qubit gate by preparing only two states and performing four measurements, reducing significantly the complexity as compared to standard process tomography techniques.
We demonstrate its applicability to a system consisting of a nitrogen-vacancy (NV) center in diamond~\cite{doherty2013nitrogen} coupled to a \nuc nuclear spin and illustrate how it can be used to optimize a two-qubit gate.

Thanks to their long coherence times at room temperature~\cite{Hanson2006,Balasubramanian2009,Maurer2012}, NV centers offer a versatile platform for quantum sensing~\cite{Schirhagl2014, glenn2018high, spohn2025quantum, vetter2022zero, Rembold2020}, quantum computing~\cite{Wrachtrup2006, neumann, van2012decoherence, Joas2024}, and quantum simulation~\cite{vetter2025room, choi2017observation}. 
The electron spin of the vacancy and the surrounding \nuc spins allow for the realization of a fully-connected spin register~\cite{Neumann2010, weber2010quantum, neumann, Chen2020, Unden2016, liu2018quantum}, with optical (even single-shot) readout~\cite{Neumann2010a, Joliffe2024}. QOC has been applied to optimize gates in NV centers both in open-loop and in closed-loop fashion~\cite{Scheuer2014, Dolde2014,Waldherr2014, Said2009,Frank2017,Rembold2020,Chen2020,Oshnik2022,Rossignolo2023,Vetter2024,Wang2025} but so far no two-qubit gates have been optimized in the closed-loop mode.

In this article we investigate a CNOT gate scheme similar to that studied by Ref.~\cite{Said2009}, but drive the gate exclusively via fast microwave control on the NV center's electron spin. We first optimize the gate in open-loop mode and then simulate a closed-loop calibration of the control pulse to a collection of sample systems with unknown parameters using a benchmarking protocol tailored to the platform. We demonstrate the applicability of the protocol and identify the key parameters for successful calibration.

The article is organized as follows. In Sec.~\ref{sec: System} we introduce the quantum system that we exploit for our simulations. In Sec.~\ref{sec: Optimal Control and Fidelity} we introduce our target gate and the relevant optimization metrics, namely figures of merit (FoM), for QOC. We illustrate and discuss our simulations results in Sec.~\ref{sec: results}. Finally, we summarize our findings and discuss future directions in Sec.~\ref{sec: conclusions}.

\section{System}\label{sec: System}

\begin{figure}
    \centering
    \includegraphics[width=\linewidth]{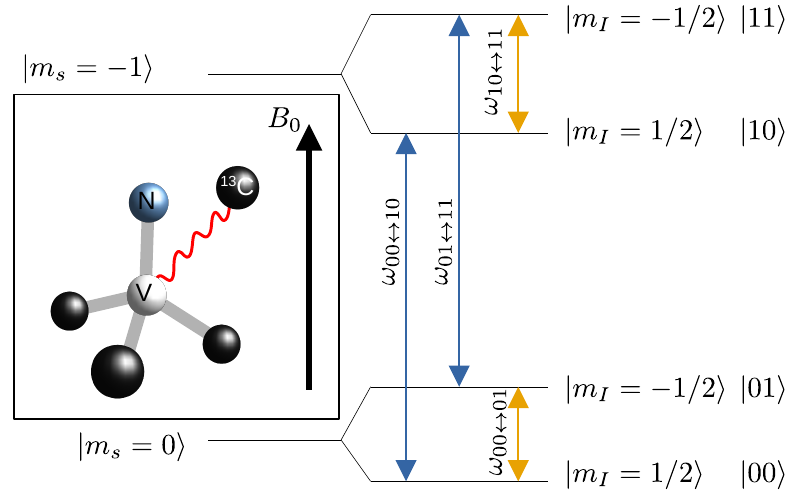}
    \caption{\textbf{System level scheme}. An NV center in the diamond lattice is coupled to a nearby \nuc spin, with the magnetic field aligned along the defect’s symmetry axis (structure in the inset). The two Zeeman levels $m_s=0$ and $m_s=-1$ on the electron and the two nuclear spin states $m_I=1/2$ and $m_I=-1/2$ form the computational basis states of the two-qubit system. Microwave (blue) and radio-frequency (yellow) pulses can drive the electronic transitions $\omega_{00\leftrightarrow10}$ and $\omega_{01\leftrightarrow11}$, and nuclear transitions $\omega_{00\leftrightarrow01}$ and $\omega_{10\leftrightarrow 11}$, respectively.}
    \label{fig:levelscheme}
\end{figure}

As a realistic framework for our method, we select a single NV center coupled to a \nuc nuclear spin (see inset in Fig.~\ref{fig:levelscheme}). Such a two-qubit system could be part of a larger spin register or directly used, e.g., for sensing with a memory qubit~\cite{Unden2016} or for light-matter interfaces~\cite{pompili21realization,kalb17entanglement,Stas2022}.
By assuming a magnetic bias field of $B_0=600\,\text{G}$ which is aligned with the NV center's symmetry axis (a common and widely used magnetic field configuration), the hyperfine interaction between the NV center’s excited state and its intrinsic nitrogen nuclear spin enables polarization of the latter via polarization transfer from the NV's electron spin. 
Consequently, the nitrogen spin is considered to be fully polarized, allowing us to neglect its interaction.
A microwave field $B_1(t)$ is used to drive the system.
Setting $\hbar=1$, the system's Hamiltonian is given by
\begin{eqnarray}
        H = D \left(S_{z}^2-\frac{2}{3}\right)+\gamma_e B_0 S_{z}+\gamma_e B_1(t) S_x\nonumber\\
    +\vec{S}^{\,\mathsf{T}}  A \vec{I}-\gamma_n B_0 I_z -\gamma_n B_1(t) I_{x}\,,
        \label{eq:system_Hamiltonian}
\end{eqnarray}
where $D=2\pi \cdot 2.87\,$GHz is the NV center's zero-field splitting, and $\gamma_e$ and $\gamma_n$ are the gyromagnetic ratio of the electron and the nuclear spin, respectively~\cite{Nizovtsev2014}. The electron spin's Zeeman levels are described by the spin-1 operator $\vec{S}$ with its components $S_x$, $S_y$ and $S_z$, and the nuclear spin by the spin-1/2 operator $\vec{I}$ with its components $I_x$, $I_y$ and $I_z$. They interact via the hyperfine interaction described by the hyperfine tensor $A$. Under the secular approximation~\cite{hall2014analytic}, the hyperfine coupling becomes
\begin{eqnarray}\label{eqn: spin approx}
    H_{\text{hf}}=\vec{S}^{\,\mathsf{T}}  A \vec{I}\approx \sum_{i=x,y,z} A_{zi}S_z I_i\,.
\end{eqnarray}
We consider a \nuc nuclear spin from the G-family~\cite{Nizovtsev2014} with $A_{zz}=2\pi\cdot2.281\,$MHz, $A_{zx}=A_{zy}=2\pi\cdot 0.240\,$MHz.
For convenience, we rotate the coordinate frame such that $A_{zx}=2\pi\cdot 0.339\,$MHz and $A_{zy}=0$.
The Zeeman interaction lifts the degeneracy of the $\vert{m_s=\pm1}\rangle$ states of the NV's electron spin and we thus choose the $\vert{m_s=0}\rangle$ and $\vert{m_s=-1}\rangle$ states as our effective qubit.
For the nuclear spin the qubit states are defined as $\vert{m_I=\pm1/2}\rangle$. 
The resulting computational basis states $\ket{x}$ for $x\in\mathcal{B}:=\pg{00,01,10,11}$ are illustrated in Fig.~\ref{fig:levelscheme}.

The drift Hamiltonian for this system is defined as
\begin{equation}\label{eqn: drift hamiltonian}
H_\text{d}=\mathrm{diag}\bigg(\omega_{00},\omega_{01},\omega_{10},\omega_{11}\bigg),
\end{equation}
where $\omega_{00}=-\frac{2}{3}D -\frac{\gamma_n B_0}{2}$, $\omega_{01}=-\frac{2}{3}D +\frac{\gamma_n B_0}{2}$, $\omega_{10}= \frac{1}{3}D -\gamma_e B_0-\frac{\gamma_n B_0}{2}$, $\omega_{11}=\frac{1}{3}D -\gamma_e B_0 +\frac{\gamma_n B_0}{2}$.
By assuming that the dynamics always stay in the computational subspace and including the control Hamiltonian ${H}_{\text{c}}$ (details in Appendix~\ref{sec: hamiltonians derivation}), the complete Hamiltonian of the system defined in a rotating frame with respect to $H_\text{d}$ becomes 
\begin{equation}\label{eqn: full system hamiltonian rot frame main}
    \tilde{H} = \tilde{H}_{\text{hf}} + \tilde{H}_\text{c},
\end{equation}
where $\tilde{H}_{\text{hf}}$ and $\tilde{H}_{\text{c}}$ denote the hyperfine and control Hamiltonians in this frame, respectively. In Appendix~\ref{sec: hamiltonians derivation} we provide the full derivation of the final expressions. Neglecting the rapidly oscillating terms we find
\begin{equation}\label{eqn: tilde_H_hf}
    \tilde{H}_{\text{hf}}=\begin{pmatrix} 
 0&0&0&0\\
 0&0&0&0\\
 0&0&-\frac{A_{zz}}{2}&-\frac{A_{zx}-iA_{zy}}{2}e^{-i\omega_nt}\\
 0&0&-\frac{A_{zx}+iA_{zy}}{2}e^{i\omega_nt}&\frac{A_{zz}}{2}
 \end{pmatrix} \, , 
 \end{equation}
 with $\omega_n=\gamma_n B_0=2\pi\cdot 642\,$kHz, and
 \begin{equation}
\tilde{H}_{\text{c}}= \begin{pmatrix}
    0&\Omega_{\text{rf,c}}(t)&\Omega_{\text{mw,c}}(t)&0\\
    \Omega_{\text{rf,c}}^{*}(t)&0&0&\Omega_{\text{mw,c}}(t)\\
    \Omega_{\text{mw,c}}^{*}(t)&0&0&\Omega_{\text{rf,c}}(t)\\
    0&\Omega_{\text{mw,c}}^{*}(t)&\Omega_{\text{rf,c}}^{*}(t)&0
\end{pmatrix}\, .
\label{eqn: H control tilde final}
\end{equation}
The full control pulses at microwave/radio-frequency transitions are described as
\begin{eqnarray}\label{eqn: def mw control}
    \Omega_{\text{mw,c}}(t)=\frac{\Omega_{\text{mw}}(t)e^{-i\phi_{\text{mw}}}}{2}e^{-i(\omega_e-\omega_{\text{mw}}) t}\,,\\
    \Omega_{\text{rf,c}}(t)=\frac{\Omega_{\text{rf}}(t)e^{-i\phi_{\text{rf}}}}{2}e^{-i(\omega_n-\omega_{\text{rf}}) t}\,,
\label{eqn: def rf control}
\end{eqnarray}
where $\omega_e=D-\gamma_e B_0$.
From Eqs.~\eqref{eqn: def mw control}-\eqref{eqn: def rf control}, control pulses $\Omega_{\text{k,c}}(t)$ for $\text{k}\in\{\text{mw}, \text{rf}\}$ are characterized by two scalar parameters (phase $\phi_{\text{k}}$ and carrier frequency $\omega_{\text{k}}$) and by the real temporal profile of the slowly-varying amplitude $\Omega_{\text{k}}(t)$. For the latter, in this work we consider the realistic experimental constraints $\abs{\Omega_{\text{mw}}(t)}\leq 2\pi\cdot 30\,$MHz for the microwave and $\abs{\Omega_{\text{rf}}(t)}\leq 2\pi\cdot 40\,$kHz for the radio-frequency~\cite{Joas2024,vetter2025room}. 

We note that the form of our full Hamiltonian is similar to the one in Ref.~\cite{Said2009}, but crucially we do not assume to have selective addressing of the four transitions. If we choose, e.g., to drive the transition $|00\rangle\leftrightarrow |01\rangle$ on resonance, our Hamiltonian also captures the effect of the off-resonant driving that such a pulse induces on the transition $|10\rangle\leftrightarrow |11\rangle$. On the other hand, the effects of the microwave control on the nuclear spin and of the radio-frequency on the electron spin are neglected, as they are far detuned.

Due to hyperfine coupling, the energy levels of the four basis states can be determined by diagonalizing the stationary Hamiltonian $H_\text{d} + H_{\text{hf}}$ (see Eq.~\ref{eqn: H_cs} in Appendix~\ref{sec: hamiltonians derivation}) or, equivalently, $H_\text{d} + \tilde{H}_{\text{hf}}$ (see Eq.~\ref{eqn: tilde_H_hf}). In doing so, one finds that the eigenenergies of the system are
\begin{subequations}
\begin{gather}
\tilde\omega_{00}=\omega_{00}, \\
\tilde\omega_{01}=\omega_{01}, \\
\tilde\omega_{10}=\frac{1}{3}D-\gamma_e B_0-\frac{1}{2}\sqrt{(\omega_n+A_{zz})^2+A_{zx}^2+A_{zy}^2}\,, \\
\tilde\omega_{11}=\frac{1}{3}D-\gamma_e B_0+\frac{1}{2}\sqrt{(\omega_n+A_{zz})^2+A_{zx}^2+A_{zy}^2}\,.
\end{gather}
\end{subequations}
Differences between these energies determine the transition frequencies among the four computational states.

\section{Optimal Control and Fidelity}\label{sec: Optimal Control and Fidelity}

In this section we discuss the optimization goal and the metrics of our analysis, introducing the gate fidelity for open-loop optimization as well as our approach to evaluate the FoM for closed-loop optimization in an efficient and practical way, including state preparation and readout. While we tailor our analysis to a specific gate, adaptions of the state preparation and measurement (SPAM) gates allow to optimize also other two-qubit gates, and the method can, in principle, be extended to three-qubit gates.

\subsection{Figures of Merit for Open- and Closed-Loop Optimization}\label{sec: Figures of Merit}
Our aim is to optimize a \cnot gate where we flip the NV center (first qubit) conditionally on the nuclear spin (second qubit).
The usual method involves applying a weak $\pi$-pulse on the NV center, which selectively addresses the hyperfine transition $|00\rangle\leftrightarrow |10\rangle$ or $|01\rangle\leftrightarrow |11\rangle$. Under perfect conditions, this leads to
\begin{eqnarray}\label{eq:target_gate}
U_{\text{targ}}=\begin{pmatrix}
    1&0&0&0\\
    0&0&0&-i\\
    0&0&1&0\\
    0&-i&0&0
\end{pmatrix}\ ,
\end{eqnarray}
which differs from the standard real-valued matrix by including a relative phase, altering some common circuit compositions \cite{barenco1995elementary, motzoi2017linear}. The gate $U_{\text{targ}}$ is the target gate of our optimization. Under the unitary time evolution of the full system $U(T)$, we can examine the fidelity at final time $T$ as \cite{Mueller2022}
\begin{eqnarray}\label{eq:fidelity}
    F_{\text{sm}}=\frac{1}{16}\left|\Tr\pt{U_{\text{targ}}^\dagger U(T)}\right|^2\,.
\end{eqnarray}
If we additionally consider a (potentially dissipative) time evolution described by the completely positive trace-preserving map $\Phi$ at time $T$, we need a total of up to 256 measurements for a full quantum process tomography of $\Phi$~\cite{OBrien2004quantum}. While this approach is already numerically expensive for open-loop optimizations, its adoption becomes extremely challenging in the closed-loop case.

A more efficient approach relies on studying the time evolution of three specific probe states~\cite{Goerz2014}, enabling the definition of an alternative and more sustainable figure of merit:
\begin{equation}\label{eq:FJ}
    F_\text{J}=\sum_{j=1}^{3}\Tr\pt{U_{\text{targ}}\rho_j U_{\text{targ}}^\dagger \Phi(\rho_j)}\,,
\end{equation}
where $\rho_1$ is a state with all different diagonal entries and zero coherences, $\rho_2$ is a pure state which is non-orthogonal to any basis state
and $\rho_3$ is an additional state to measure leakage. 
We remark that, as shown in Ref.~\cite{Goerz2014}, $F_\text{J}$ is not equivalent to the gate fidelity $F_{\text{sm}}$ in Eq.~\eqref{eq:fidelity}, but for a close-to-unitary time evolution the maximization of $F_\text{J}$ does correspond to that of $F_{\text{sm}}$ \cite{Reich2013,Romer2025}.

For simplicity, in this work we consider the case in which the system does not leave the qubit subspace, thus focusing only on the evolution of the first two states. 
In detail, we set our target probe states for the optimization to
\begin{gather}\label{eq:rho1_th}
\rho_1 = \text{diag}(0.23, 0.56, 0.06, 0.15)\,, \\
\rho_2=\frac{1}{4}\begin{pmatrix}
        1&1&1&1\\
        1&1&1&1\\
        1&1&1&1\\
        1&1&1&1
    \end{pmatrix} .\label{eq:rho2_th}
\end{gather}
The choice of $\rho_1$ has been made by maximizing the minimum difference among diagonal terms when performing a single state preparation pulse on the nuclear spin and one on the electron spin (see Eq.~\eqref{eq:rho1-parameters} and discussion in Sec.~\ref{sec: state prep and readout}).

Finally, to evaluate the quality of state preparations we adopt the canonical fidelity between two quantum states $\rho$ and $\sigma$, defined as
\begin{eqnarray}\label{eqn:state fidelity}
    F\pt{\rho,\sigma} = \left[\, \mathrm{Tr}\!\left( 
\sqrt{ \sqrt{\rho}\, \sigma \, \sqrt{\rho} }
\right) \right]^2.
\end{eqnarray}

\subsection{Practical approach to $F_\text{J}$ evaluation}\label{sec: stochastic approach}
Typically, quantum systems are controlled via unitary operations. Assuming that the system is initialized in a basis state, e.g. $\rho_0=\ketbra{00}$, this does not pose any issues in the preparation of the pure state $\rho_2$ in Eq.~\eqref{eq:rho2_th}, since the two pure states are connected via unitary transformations. However, it is challenging to prepare a controlled mixture like $\rho_1$ (Eq.~\eqref{eq:rho1_th}), as this requires a non-unitary process.  On the other hand, starting from $\rho_0$, it is relatively easy to prepare a pure state $\tilde{\rho}_1$ that satisfies
\begin{equation}\label{eq: equal diagonal for rho1, rho1tilde}
    \mathrm{diag}(\tilde{\rho}_1) = \mathrm{diag}({\rho}_1)
\end{equation}
via unitary transformations.

Note that infinitely many such states exist, as Eq.~\eqref{eq: equal diagonal for rho1, rho1tilde} holds for states with arbitrary off-diagonal terms. More formally, suppose that one of such states has been realized 
by performing $N$ (either microwave or radio-frequency) controlled pulses according to Eq.~\eqref{eqn: def mw control} on the initialized two-qubit system, and let $\pg{\phi_1,\ldots,\phi_N}$ denote the ordered set of phases of these pulses, which determine for each of them the rotation axis on the $x-y$ plane of the spin.
Now, let us consider the set $\mathcal{S}$ defined by all possible combinations of $N$ phases, where for each index $j=1,\ldots,N$ we either pick the original value of $\phi_j$ or its complementary $\phi_j+\pi$. We obtain a set of $2^N$ $N$-tuples:
$\mathcal{S} = \pg{\phi_1, \phi_1+\pi} \times \pg{\phi_2, \phi_2+\pi} \times \ldots \times \pg{\phi_N, \phi_N+\pi}$. For each $N$-tuple $s\in\mathcal{S}$, let $\tilde{\rho}_1^{s}$ denote the state obtained by applying the same controls and setting these phases. We observe that
\begin{equation}
    \sum_{s\in\mathcal{S}} \mel{x}{\tilde{\rho}_1^{s}}{y} \equiv 
    \Big\langle{\mel{x}{\tilde{\rho}_1^{s}}{y}}\, \Big\rangle_{s\in\mathcal{S}} = \mel{x}{\rho_1}{y}
    = \delta_{xy}\lambda_{y},
\end{equation}
where $x,y \in\mathcal{B}$, $\delta_{xy}$ is the Kronecker delta and $\lambda_{y} := \expval{\rho_1}{y}$. Intuitively, this is because for every rotation along a given axis of a spin, we are also considering the complementary back-rotation along the same axis in another tuple of phases, and thus when considering all possible realizations these effects average out.

As a result, the average of all these states yields an effective representation of $\rho_1$:
\begin{equation}
\expval{\tilde{\rho}_1^{s}}_{s\in\mathcal{S}} = \rho_1.
\end{equation}
We note that the contribution of $\rho_1$ to the fidelity $F_J$ in Eq.~\eqref{eq:FJ} is linear in $\rho_1$, implying that 
\begin{align}\nonumber
\Tr\pt{U_{\text{targ}}\rho_1 U_{\text{targ}}^\dagger \Phi(\rho_1)} &=
\Tr\pt{U_{\text{targ}}\rho_1 U_{\text{targ}}^\dagger \Phi(\expval{\tilde{\rho}_1^{s}}_{s\in\mathcal{S}} )}  \\
&= \expval{\Tr\pt{U_{\text{targ}}\rho_1 U_{\text{targ}}^\dagger \Phi(\tilde{\rho}_1^{s})}}_{s\in\mathcal{S}}.
\label{eqn: randomized rho_1 fidelity}
\end{align}
It follows that we can compute the contribution of $\rho_1$ to $F_J$ by averaging over the score of the states $\tilde{\rho}_1^{s}$.

The above arguments hold for quantum systems of arbitrary dimension. In our two-qubit system, we prepare $\tilde{\rho}_1$ with $N=2$ pulses and thus need $4$ different combinations of the two pulse phases (see also Sec.~\ref{sec: state prep and readout}).
Instead of using pulses with deterministic complementary phases one could also sample the $N$ phases randomly out of a uniform distribution and prepare $\tilde{\rho}_1$ many times. We checked that about $10^5-10^6$ random realizations would be enough to prepare the ensemble average $\langle\tilde{\rho}_1^s\rangle\approx\rho_1$ with sufficient precision for our system.

Crucially, both choices enable an efficient evaluation of the contribution of $\rho_1$ to $F_J$. In fact, for many systems, gathering readout statistics inevitably requires to prepare, evolve and measure multiple times a certain quantum state, in order to achieve the desired readout precision. As a result, we can exploit this inherent requirement to prepare the state with different pulse phases at limited extra cost, resulting in the effective preparation of a mixed state.

We show in Sec.~\ref{sec: results} that optimizing the control pulses over the average score of $\tilde{\rho}_1^s$ indeed results as well in optimizing the value of the reference fidelity $F_{\text{sm}}$, within the limits of realistic implementation noise.

\begin{figure*}
    \centering
    \includegraphics[width=0.99\linewidth]{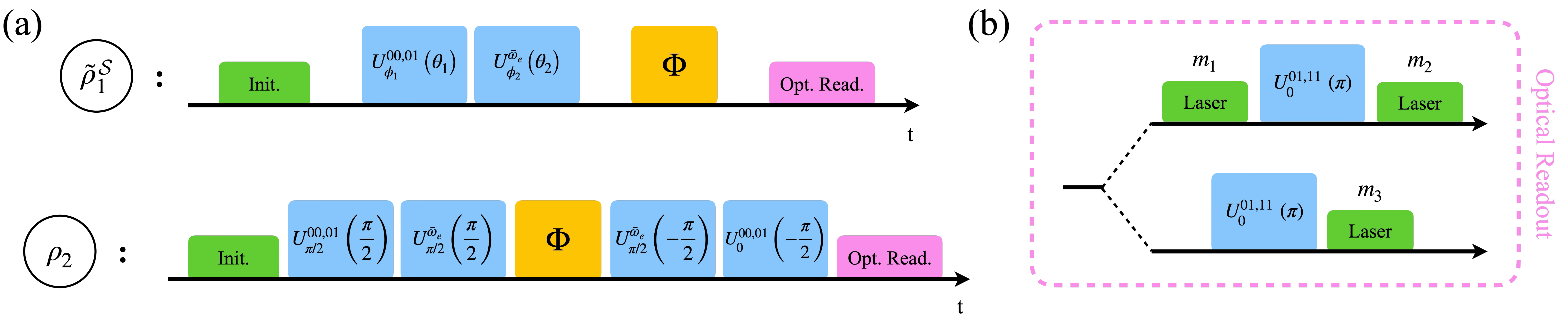}
    \caption{\textbf{Full pulse sequences for (a) state preparation and (b) optical readout.} We illustrate the full pulse sequences to experimentally compute ${F}_\text{J}$. Green pulses represent the initialization/readout action induced by a laser, while blue pulses are required to prepare correctly the probe states and measure the trial control action on them. The QOC gate itself is represented by the time evolution map $\Phi$ (yellow). Finally, the two-step readout routine (pink), further illustrated in panel (b), allows to access all the diagonal terms of the final states as described in Eq.~\eqref{eqn: OptRead conversion}. Full details of the pulse sequence can be found in Sec.~\ref{sec: state prep and readout}.}
    \label{fig:Meas_scheme_label}
\end{figure*}

\subsection{Pulse Parametrization and Optimization Algorithm}\label{sec: Pulse Parametrisation and optimization Algorithm}

To generate the \cnot gate (i.e., the imperfect transformation $\Phi$ introduced in Sec.~\ref{sec: Figures of Merit}) we apply and optimize a time-modulated microwave pulse $\Omega_{\text{mw,c}}(t)$ as introduced in Eq.~\eqref{eqn: def mw control}.
As we drive the optimal-control gate only via microwave control, we set $\Omega_{\text{rf,c}}(t)\equiv0$ (constant radio-frequency controls are instead used for some of the SPAM gates, as indicated in Sec.~\ref{sec: state prep and readout}). We further parameterize the microwave control amplitude as follows:
\begin{eqnarray}\label{eqn: Omega factorisation}\Omega_{\text{mw}}(t)=\Omega_{\text{mw,}0} \, g(t) \, f(t)\,.
\end{eqnarray}
Here, $\Omega_{\text{mw,}0}$ is the constant amplitude of the pulse, while $g(t)$ is the envelope function
\begin{equation}\label{eq:myfunc}
g(t) = \left\{ 
\begin{array}{ll}
\frac{1}{\alpha} \!\left(
e^{-\frac{(t-t_r)^2}{2\sigma^2}}
- e^{-\frac{t_r^2}{2\sigma^2}}
\right), & t \le t_r, \nonumber \\
1, & t_r < t < T - t_r, \nonumber \\
\frac{1}{\alpha} \!\left(e^{-\frac{(t - T + t_r)^2}{2\sigma^2}}
- e^{-\frac{t_r^2}{2\sigma^2}}
\right), & t \ge T - t_r, 
\end{array}
\right.
\end{equation}
where $T$ denotes the total pulse time, $t_r=0.1\cdot T$ is the rise rime, $\sigma = {t_r}/{4}$ and $\alpha = 1 - \exp\!\left(-\frac{t_r^2}{2\sigma^2}\right)$.
Finally, $f(t)$ is the pulse modulation function 
which is expanded into a set of basis functions according to the dCRAB algorithm as implemented in QuOCS~\cite{Rossignolo2023}:
\begin{eqnarray}
f(t)=\sum_{j=1}^{N_s}\sum_{k=1}^{N_c} a_k^{(j)}\sin(\omega_k^{(j)} t)+b_k^{(j)}\cos(\omega_k^{(j)} t)\,,
\end{eqnarray}
where $N_s$ is the number of super-iterations of dCRAB, $N_c$ is the number of basis functions per super-iteration and $\omega_i^{(j)}$ are the random CRAB frequencies. These frequencies are chosen randomly from uniform distributions corresponding to the intervals $[2\pi (k-1) /T,2\pi k /T]$.

We then optimize for each super-iteration $j$ the coefficients $a_{k}^{(j)}$ and $b_{k}^{(j)}$, together with the constant phase $\phi_{\text{mw}}$, the carrier frequency $\omega_{\text{mw}}$ (see Eq.~\eqref{eqn: def mw control}) and pulse duration $T$ via the Nelder-Mead algorithm. 
All optimizations are carried out with the dCRAB method of QuOCS~\cite{Rossignolo2023}.

\section{Results}\label{sec: results}

In our analysis, we first perform an open-loop optimization and subsequently adapt the results to a set of twenty sample systems by simulating a closed-loop optimization using our practical approach described in Sec.~\ref{sec: stochastic approach}. 
These NV–\nuc systems are generated by drawing random values of $A_{zx}$ and $A_{zz}$ from Gaussian distributions centered on the nominal parameters given in Sec.~\ref{sec: System}, with standard deviations of $5\%$ of the nominal values. This resembles the experimental uncertainty that one can expect to encounter when measuring the parameters of the quantum system.
We calibrate the control pulse in a closed-loop manner by adjusting only a small number of pulse parameters: this demonstrates the applicability of our method under conditions that closely resemble an experiment, showing how a numerical optimization can be efficiently adapted to a real setup.

Before discussing the actual optimization results, we introduce the realistic state preparation and readout process we simulate.

\subsection{State Preparation and Readout}\label{sec: state prep and readout}

We assume that the system can be initialized in the state $\rho_0=\ketbra{00}$: the NV's electron spin can be initialized by optical excitation (via a green laser pulse) into the excited state and by spin-selective shelving into a singlet state, while the nuclear spin can be initialized by polarization transfer from the electron spin. We further assume that the same laser pulse that is used for initialization is also used to readout the population of the electron state $\ket{m_s=0}$.
The latter corresponds to a measurement of the sum of the first two diagonal elements of a generic density matrix $\rho$, i.e., $\mel{00}{\rho}{00}+\mel{01}{\rho}{01}$. All relevant states must be prepared from $\rho_0$ by additional microwave and radio-frequency pulses.
Similarly, all readout operations are based on the population measurement of the electron state $\ket{m_s=0}$, combined with additional pulses. The errors from these pulses are considered in our simulations. 

Here, we show how to perform the necessary transformations by a series of constant pulses. The pulse scheme for state preparation and readout is illustrated in Fig.~\ref{fig:Meas_scheme_label}.
For all sequences, a crucial challenge of the system is the dephasing of the electron spin state when it is in a superposition. Therefore, we design the sequences such that this superposition only occurs for a minimal time: both the QOC gate itself and the SPAM gates on the electron have to be fast, while the slow rf gates on the nuclear spin are applied while the electron is not in a superposition state. The total time spent in the superposition state is then about 1\,\textmu s (well below typical dephasing times, especially for isotopically diluted samples~\cite{Balasubramanian2009,Maurer2012}) and we do not explicitly include the dephasing in our simulations.

\textbf{Pulse description:} We describe the gates for state preparations and readout by means of operators in the form $U^{x,y}_{\phi}\pt{\theta}$. This notation refers to the control Hamiltonian in Eq.~\eqref{eqn: H control tilde final} driven at a frequency $\omega_{\text{k}} = \tilde\omega_{x\leftrightarrow y}$, where $x,y\in\mathcal{B}$ and k\,=\,mw (k\,=\,rf) for controls addressing the electron (nuclear) spin. For simplicity, for state preparation and readout we consider constant pulses whose duration is specified accordingly in the text, while their amplitude is calibrated to provide a rotation of angle $\theta$ around the designated axis of the spin which lays in the $x-y$ plane. This axis is specified by the control phase $\phi_{\text{k}}$, where $\phi_{\text{k}}=0$ ($\phi_{\text{k}}=\pi/2$) indicates the $x$ ($y$) axis. 
Non-selective transition on the electron spin are driven at a frequency
$\bar{\omega}_{\text{e}} := (\tilde\omega_{00\leftrightarrow10} + \tilde\omega_{01\leftrightarrow11})/2$.

\textbf{Note on pulse duration:}
Due to the choice of the reference system, in our simulations, pulses are more accurate \cite{Zeuch2020} when their length $T$ is a multiple of the characteristic duration of the hyperfine interaction given by
\begin{equation}\label{eqn: T_hf}
    T_{\text{hf}} :=  2\pi \cdot \frac{2}{\sqrt{A_{zx}^2 + A_{zz}^2}} \simeq 0.87\, \text{\textmu s}.
\end{equation}
We thus choose the durations of all pulses as a multiple of $T_{\text{hf}}$, unless they are required to be fast (which is, for example, the case for the SPAM gates that bring the electron spin in a superposition state).

\textbf{Preparation of $\tilde{\rho}_1^{s}$\,:} 
To make the state preparation readily implementable in an experiment, we consider a state preparation for the states $\tilde{\rho}_1^{s}$ as introduced in Sec.~\ref{sec: stochastic approach}, which consists only of two constant pulses: a rotation on the nuclear spin followed by one on the electron spin (see Fig.~\ref{fig:Meas_scheme_label}). Ref.~\cite{Goerz2014} discusses the importance of the diagonal entries of this state being different, since similar entries could direct the optimization towards local minima and wrong gates. Therefore, we optimize this state preparation stage looking for the rotation angles $\theta_1, \, \theta_2$ which maximize distinguishability of the diagonal entries, namely
\begin{eqnarray}\label{eq:rho1-parameters}
    \theta_1, \theta_2 = \text{argmax}\pq{\min_{\substack{x,y\in\mathcal{B} \\ x\neq y}} \abs{\expval{U_{\theta}\rho_0\,U_{\theta}^{\dag}}{x} - \expval{U_{\theta}\rho_0\,U_{\theta}^{\dag}}{y}}}\,, \nonumber \\
    U_{\theta} := U^{\Tilde{\omega}_e}_{0}\pt{\theta_2} U^{00,01}_{0}\pt{\theta_1} \,.\qquad\qquad
\end{eqnarray}
The optimal rotation angles are found to be $\theta_1 = 37\pi/58$ and $\theta_1 = 3\pi/10$, and the diagonal of the corresponding evolved state is reported in Eq.~\eqref{eq:rho1_th}.

To make off-diagonal terms vanish, we consider four phase tuples ($\mathcal{S} = \pg{\pt{0,0},\pt{0,\pi},\pt{\pi,0},\pt{\pi,\pi}}$) and for each setting $s=\pt{\phi_1,\phi_2}\in\mathcal{S}$ we prepare the corresponding state $\tilde{\rho}_1^{s}$ as illustrated in Fig.~\ref{fig:Meas_scheme_label}. 
We first apply a slow rf pulse of duration of about $50 \, \text{\textmu s}$ (more precisely, $T = 57\cdot T_{\text{hf}} = 49.5 \, \text{\textmu s}$, the closest integer multiple of $T_{\text{hf}}$). This slow rf pulse generates a selective transition. The rf pulse is followed by a fast mw pulse on the NV's electron spin ($T=10 \, \text{ns}$) to limit the time the electron spends in a superposition state. The corresponding Rabi frequencies read $\Omega_{\text{rf}} = 2\pi\cdot6.5\,$kHz for the first and $\Omega_{\text{mw}} = 2\pi\cdot15\,$kHz for the second pulse, respectively.

The state preparation fidelity between the average of the states $\tilde{\rho}_1^{s}$ and the target state $\rho_1$ in Eq.~\eqref{eq:rho1_th} is computed as per Eq.~\eqref{eqn:state fidelity}, yielding $F\pt{\rho_1,\expval{\rho_1^{s}}_{s\in\mathcal{S}}} = 99.9993\%$.

\begin{figure*}
    \centering
    \includegraphics[width=0.99\linewidth]{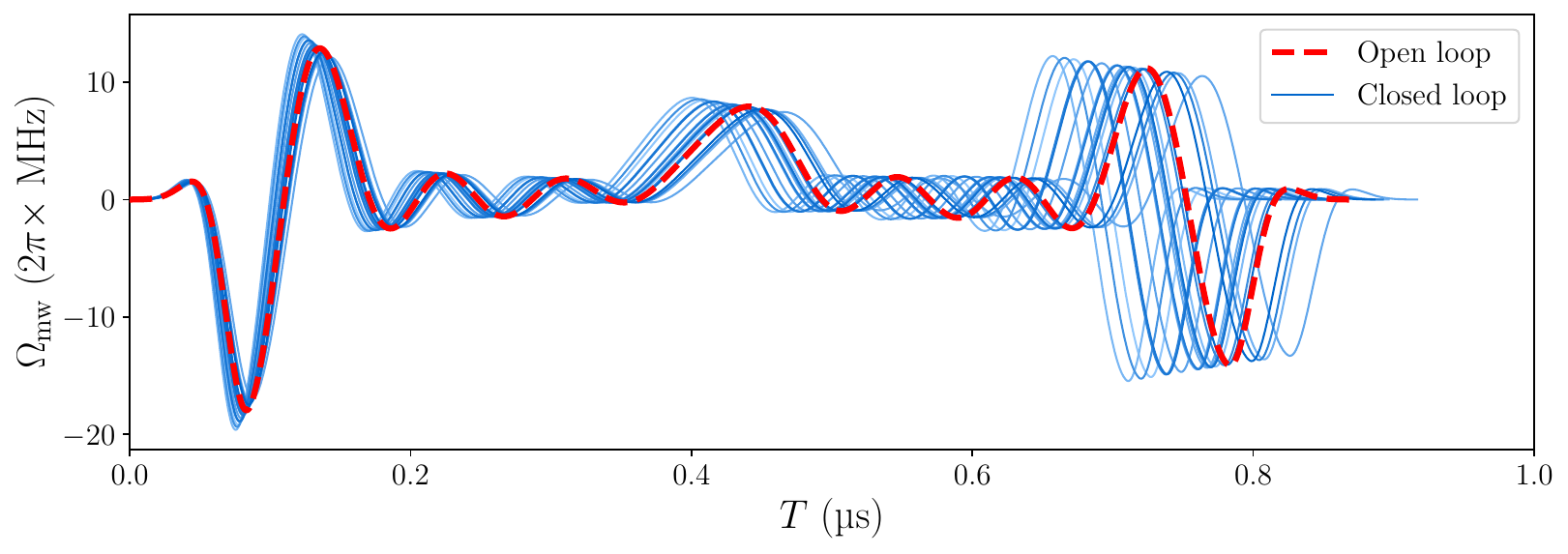}
    \caption{\textbf{Optimized pulse shapes.} The dashed red line denotes the best microwave pulse shape obtained from the open-loop optimization, while the other twenty lines denote the rescaled pulses (both in amplitude and time) found by closed-loop optimization of the sample systems. Their performance is summarized in the first column of Fig.~\ref{fig:AllResultsSummary}.}
    \label{fig:manyFoMs}
\end{figure*}

\textbf{Preparation of $\rho_2$\,:} 
To prepare $\rho_2$ (second row of Fig.~\ref{fig:Meas_scheme_label}) we first apply a slow $\pi/2$ rotation on the nuclear spin around the $y$ axis ($T = 49.5 \, \text{\textmu s}$), followed by a fast $\pi/2$ rotation on the electron spin around the $y$ axis ($T = 10$\,ns).
The fidelity between the state $\rho_2^{\text{prep}}$ prepared with these pulses and the ideal one $\rho_2$ in Eq.~\eqref{eq:rho2_th} is
$F(\rho_2,\rho_2^{\text{prep}}) = 99.93\%$.

\textbf{Readout:} We tailor the readout procedure to be convenient for the two probe states. For the first term of Eq.~\eqref{eq:FJ}, we note that the action of the target gate $U_{\text{targ}}$ on $\rho_1$ simply swaps the second and fourth diagonal entry, yielding a state $\rho_1' = \text{diag}\pt{\lambda_{00},\lambda_{11},\lambda_{10},\lambda_{01}}$. Therefore
\begin{equation}
    \Tr\pt{ U_{\text{targ}}\rho_1 U_{\text{targ}}^\dagger \Phi(\rho_1)} =
    \sum_{x\in\mathcal{B}} \expval{\rho_1'}{x} \expval{\Phi(\rho_1)}{x}\,.
\end{equation}

We thus have to measure only the diagonal entries of $\Phi(\rho_1)$. This can be achieved by combining a selective $\pi$-pulse on the electron spin with basic population measurements, as illustrated in Fig.~\ref{fig:Meas_scheme_label}. Such a selective $\pi$-pulse needs to swap the population of the $\ket{01}$ and $\ket{11}$ states, which is almost exactly the action of the target gate \cnot itself. However, note that rather than the full unitary, here it is sufficient to maximize the contrast of this conditional flip given by
\begin{gather}
    C := \abs{\sum_{x\in{\pg{00,01}}} \expval{\rho_{00}'}{x}  - \sum_{x\in{\pg{00,01}}} \expval{\rho_{01}'}{x}}
    \,, \nonumber  \\
    \rho_{x}' := U^{01,11}_0\pt{\pi}\ketbra{x}\pt{U^{01,11}_0\pt{\pi}}^{\dag} \text{ for } x\in{\pg{00,01}} \,.
\end{gather}
We find that a non-optimized constant pulse on resonance with the $\tilde\omega_{01\leftrightarrow11}$ transition (which yields a suboptimal \cnot gate) suffices in enabling high contrast measurement of the diagonal entries. In particular, we consider an electron $\pi$-pulse $U^{01,11}_0\pt{\pi}$ of duration $T=2\cdot T_{\text{hf}} = 1.74 \, \text{\textmu s}$ ($\Omega_{\text{mw}} \approx 2\pi\cdot 300\,$kHz) whose reference fidelity with the target gate is $F_{\text{sm}} = 99.3\%$, but that achieves a contrast $C = 99.96\%$. This contrast can also be calibrated directly in an experiment without full knowledge of the action of the rectangular pulse on the nuclear spin (e.g., it is possible to achieve perfect contrast even if the pulse flips the nuclear spin as well \cite{Gundlapalli2025}, or the phase relations between the action on different basis states differ from the target gate in Eq~\eqref{eq:target_gate}).
 
By preparing the time-evolved state $\Phi(\rho_1)$ twice and performing two different procedures for readout, we can retrieve a total of three values of the population of the electron state $\ket{m_s = 0}$, namely the terms
$m_1,m_2$ and $m_3$ in Fig.~\ref{fig:Meas_scheme_label}(b). From these we can reconstruct the diagonal terms of a generic state $\rho$ as
\begin{equation}\label{eqn: OptRead conversion}
   \text{diag}\left(\rho\right) = \frac{1}{2}
\begin{bmatrix}
1 & 1 & 1 & -1 \\
1 & -1 & -1 & 1 \\
-1 & 1 & -1 & 1 \\
-1 & -1 & 1 & 1
\end{bmatrix}
\begin{bmatrix}
m_1 \\
m_2 \\
m_3 \\
1
\end{bmatrix}.
\end{equation}

For the second term of Eq.~\eqref{eq:FJ}, we note that by considering $V_2=\exp(i\pi/4\sigma_y)\, \otimes\, \exp(i\pi/4\sigma_x)$ we have
\begin{equation}\label{eq:bringbackrho2}
    V_2\, U_{\text{targ}}\,\rho_2\, U_{\text{targ}}^\dagger V_2^\dagger = \rho_0,
\end{equation}
which in turn implies
\begin{align}\label{eq:measurerho2}
    \Tr\left(U_{\text{targ}}\rho_2 U_{\text{targ}}^\dagger \Phi(\rho_2)\right) &=\Tr\pt{\rho_0  V_2 \Phi(\rho_2)V_2^\dagger} 
    \nonumber \\
    &= \expval{V_2 \Phi(\rho_2)V_2^\dagger}{00}\,.
\end{align}
From Eq.~\eqref{eq:bringbackrho2}, we learn that the action of $V_2$ on the transformed state $U_{\text{targ}}\rho_2 U_{\text{targ}}^\dagger$ is that of bringing it back to the initial state. As a consequence, after applying $V_2$ to the time-evolved state $\Phi(\rho_2)$, we just have to read out the first diagonal element of its density matrix, as we can see from Eq.~\eqref{eq:measurerho2}. This can be done in the same way as for $\Phi(\rho_1)$ (see Fig.~\ref{fig:Meas_scheme_label}), i.e., preparing also $\rho_2$ twice, and measuring the time-evolved state with the two different readout procedures.

The transformation $V_2$ consists of a $(-\pi/2)_y$-pulse on the electron (executed by a fast microwave pulse over $10$\,ns) followed by a $(-\pi/2)_x$-pulse on the nuclear spin (a weak radio-frequency pulse with $T=49.5 \, \text{\textmu s}$). Note that this time we first apply the microwave control to keep the electron in a superposition state as little as possible, and thus its dephasing only plays a role starting from the fast microwave pulse. 

We remark that all these pulses are simulated by computing the corresponding unitary operators through the full system Hamiltonian (Eq.~\eqref{eqn: full system hamiltonian rot frame main}), tuning the control term. While in the definition of the pulses for the state preparation and readout we assume the transition frequencies to be known (which might require additional calibration, see discussion in Sec~\ref{sec: closed loop}), we take into account all effects in terms of finite precision of the pulses, cross-talk among transitions and overall imperfect state preparation and readout. This choice is equivalent to assuming that the pulses required for state preparation and readout have been calibrated independently. 

\subsection{Open-Loop optimization}\label{sec: Open-loop optimization}

When performing an open-loop optimization, we have access at any moment to the density matrix of the simulated evolved state. Therefore, we directly maximize the fidelity measure $F_{\text{sm}}$ introduced in Eq.~\eqref{eq:fidelity}.
We choose $N_c=10$ basis functions for each of the $N_s=15$ super-iterations and optimize $\Omega_{\text{mw}}(t)$ as described in Sec.~\ref{sec: Pulse Parametrisation and optimization Algorithm}, together with $\phi_{\text{mw}}$, $\omega_{\text{mw}}$ and the pulse duration $T$. Initial guesses are set to $0$ for phase and amplitude, to $\Tilde{\omega}_{01\leftrightarrow11}$ for $\omega_{\text{mw}}$ and to $1 \, $\textmu s for $T$. Notably, the optimal pulse duration found is $T \approx T_{\text{hf}}$, 
which corresponds to the period of the hyperfine coupling introduced in Eq.~\eqref{eqn: T_hf}.
As a result, we fine-tune the optimization over the pulse amplitude, phase and carrier frequency, while setting the fixed pulse duration to $T_{\text{hf}}$.

The best pulse shape found by the algorithm for the open-loop optimization is depicted in Fig.~\ref{fig:manyFoMs}, together with the results of the closed-loop analysis we discuss in the next section. The corresponding fidelity value is $F_{\text{sm}} = 99.97\%$.

\begin{figure}
    \centering
    \includegraphics[width=0.9\linewidth]{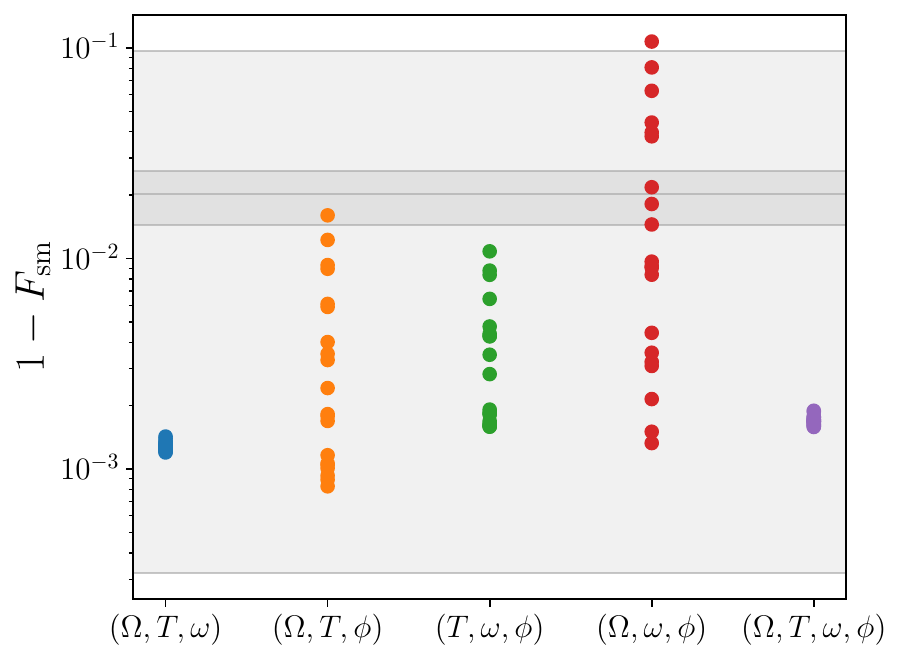}
    \caption{\textbf{Closed-loop optimization scores for different parameter choices.} Reference infidelity achieved when adapting the original open-loop solution to twenty sample systems: for each of them we optimize over all four scaling parameters and over every combination of three of them (subscripts are removed for convenience). The background describes the reference infidelity obtained by plugging directly the open-loop solution in the sample systems: top and bottom line mark respectively the highest and lowest infidelity encountered, while the central, darker region indicates the mean value and its stochastic error.}
    \label{fig:AllResultsSummary}
\end{figure}

We also analyze how stable this pulse is when it is directly applied to the twenty sample systems representing possible experimental cases. Results vary as follows: while in some exceptional case the open-loop solution performs even better on a different system, in the vast majority of cases  the performance degrades considerably. A visual representation of this can be found as the background of Fig.~\ref{fig:AllResultsSummary} (light gray shades), where the maximum, minimum and average score of this pulse over the sample systems is displayed (gray lines), together with the improved results of the closed-loop optimization we discuss in the next section.

\subsection{Closed-Loop Optimization}\label{sec: closed loop}

In this section we discuss how to calibrate the open-loop results to different sample systems. To do so, we simulate step-by-step the full workflow of the closed-loop scheme, including state preparation, application of the control pulse and optical readout, as discussed in Sec.~\ref{sec: state prep and readout}. We remark that here we do not consider readout noise, but we do take into account imperfections in the SPAM gates as all the necessary pulses are simulated through the control Hamiltonian in Eq.~\eqref{eqn: H control tilde final}.

To keep the optimization light, as required in a realistic experimental scenario, here we do not optimize the pulse shape. Instead, we start from the optimal results of the open-loop optimization discussed in the previous section, and fine-tune four scaling parameters to adapt the open-loop solution to each sample system (see Fig.~\ref{fig:manyFoMs}). These parameters are $\Omega_{\text{mw,}0}$, $\omega_{\text{mw}}$, $\phi_{\text{mw}}$ and $T$ (see Secs.~\ref{sec: Pulse Parametrisation and optimization Algorithm} and \ref{sec: Open-loop optimization}). Note that in this case optimization of $\omega_{\text{mw}}$ is required, as we do not know \textit{a priori} the value of the transition frequencies.

\begin{figure}
    \centering
    \includegraphics[width=\linewidth]{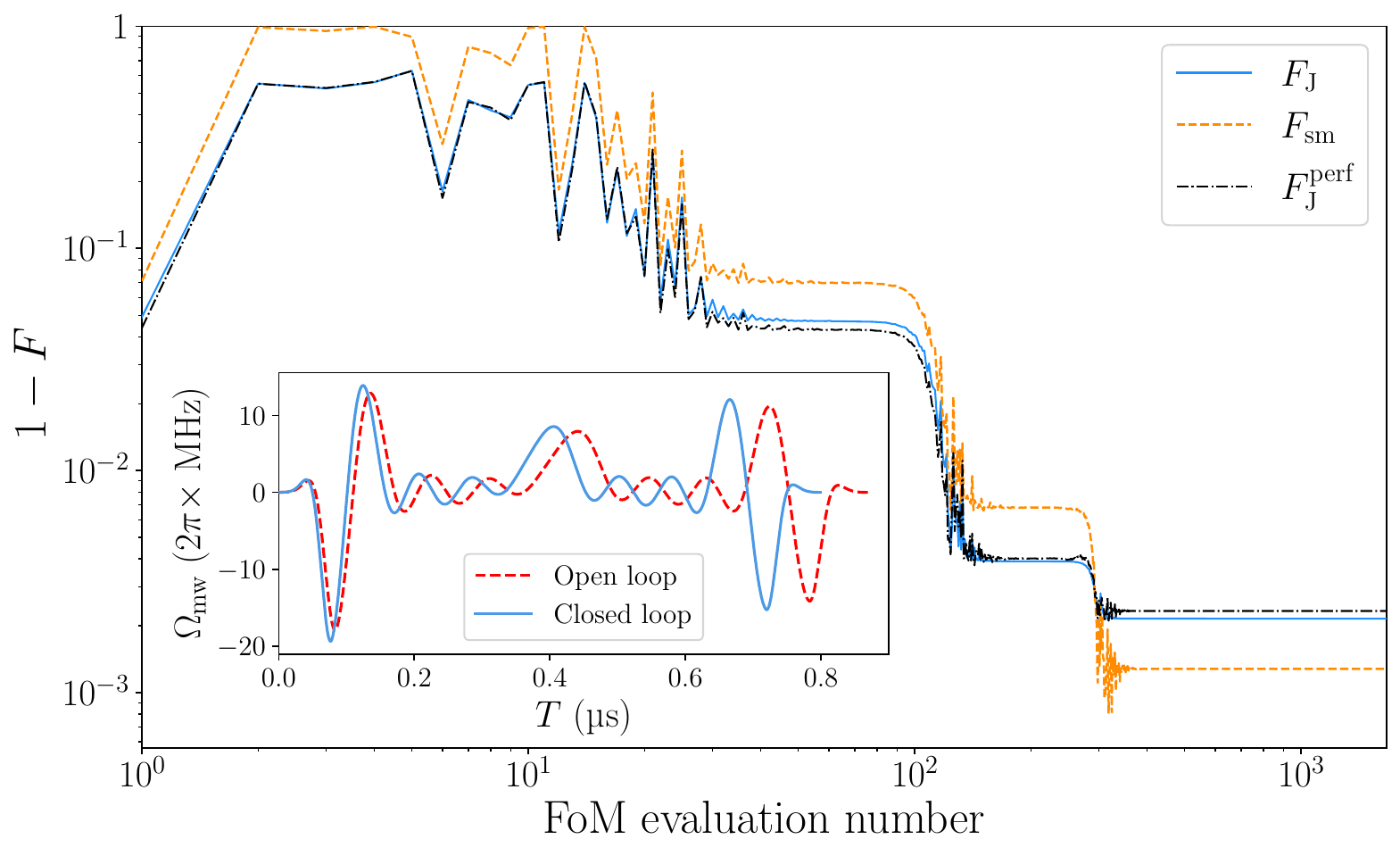}
    \caption{\textbf{Example of closed-loop optimization.} The main plot shows the optimization process, comparing the optimized infidelity (solid blue line with SPAM errors; dot-dashed black line without SPAM errors) and the reference infidelity (dashed orange line) over iterations.
    The inset shows the final rescaled closed-loop pulse (solid blue line), compared with the original open-loop solution (dashed red line).}
    \label{fig:Example_onepic}
\end{figure}

For each sample system, the closed-loop search is performed adopting the Nelder-Mead algorithm and maximizing the FoM $F_\text{J}$ in Eq.~\eqref{eq:FJ}, according to the prescriptions in Sec.~\ref{sec: state prep and readout}. 
In order to maximize efficiency, we aim to reduce the amount of parameters subject to optimization as much as possible. Therefore, after performing the aforementioned analysis for the four parameters $\pt{\Omega_{\text{mw}}, T,\omega_{\text{mw}},\phi_{\text{mw}}}$, we repeat it for each possible combination of three of them. In these analyses the non-optimized parameter is kept fixed at the best value returned by the open-loop search. 
The results are compared in Fig.~\ref{fig:AllResultsSummary}, including the open-loop case.
We underscore that Fig.~\ref{fig:AllResultsSummary} refers to the reference fidelity measure $F_{\text{sm}}$ (which is different from the one actually computed for the optimization, i.e., $F_J$) for a fair comparison of the results. 

Overall, results suggest that optimization of the pulse duration and carrier frequency is crucial (as results without calibration of these parameters display the highest dispersion). On the other hand, it looks like the phase does not play any role: when comparing the first and last column in Fig.~\ref{fig:AllResultsSummary}, we observe that adding the phase as a free parameter actually enlarges the search space and eventually drives the algorithm towards local minima, thus worsening slightly the results. We note that this behavior is also present when the simplex size for the algorithm search is too large, and the results in Fig.~\ref{fig:AllResultsSummary} are thus obtained with very small simplex sizes, keeping the search in the neighborhood of the open-loop solution.

The scalar optimization of $\Omega_{\text{mw}},\, T$ and $\omega_{\text{mw}}$ improves the open-loop control on the sample systems by over an order of magnitude on average, settling around values of $F_{\text{sm}}\simeq 99.9\%$ over about $1000$ iterations. The rescaled pulse shapes resulting from these optimizations are depicted in Fig.~\ref{fig:manyFoMs}, compared to the open-loop solution, while an example of an optimization run is illustrated in Fig.~\ref{fig:Example_onepic}. We believe that this fidelity is mainly limited by two factors: first, the imperfect SPAM gates lead to a discrepancy between the optimized FoM $F_\text{J}$ (which could potentially encounter additional initialization and readout errors in an actual experiment) and the perfect FoM $F_{\text{J}}^{\text{perf}}$ that we would obtain without SPAM errors
(compare the solid blue line and dot-dashed black line in Fig.~\ref{fig:Example_onepic}). The effect of this discrepancy could in principle be reduced by repetitive benchmarking (i.e. multiple application of the QOC gate in each measurement sequence). The second major limitation comes from our four-parameter ansatz for the closed-loop optimization, which allows us to calibrate the open-loop solution in only a small number of iterations. Indeed, while for the example case in Fig.~\ref{fig:Example_onepic} the FoM was evaluated 1666 times, there is no more visible improvement after about 400 evaluations. If more evaluations are feasible in an experiment, a full optimization of the pulse shape could lead to a better result.

In the regime we have studied, despite the realistic imperfections in the simulation of state preparation and readout, the overall minimization of $1-{{F}_{\text{J}}}$ also results in a minimization of $1-{{F}}_{\text{sm}}$ since the discrepancy coming from the SPAM errors is not very high. We note that after removing all preparation and measurement imperfections (dot-dashed black line in Fig.~\ref{fig:Example_onepic}) the behavior of the two metrics is indeed very similar, although Ref.~\cite{Goerz2014} only guaranties that the ultimate optimal solution matches, and not necessarily that the same occurs for behavior further away from the optimum.

\section{Conclusions}\label{sec: conclusions}
We have shown a pathway toward closed-loop calibration of a two-qubit gate with NV centers in diamond based on a scheme that, due to the realization of carefully chosen states, under realistic experimental conditions only requires 4 measurements in contrast with the 144 required by standard quantum process tomography, marking a significant improvement. Our scheme is tailored to a CNOT gate implementation and a specific NV centers platform, but adaptions for similar systems or locally equivalent gates seem straightforward. Further generalizations could be a topic for future work.
Moreover, we have identified for our system the key parameters of the control pulse that have to be calibrated in order to adjust an open-loop solution to the deviations between model and device. While we have explicitly studied QOC for NV centers, the main results are readily extendable to other control approaches or platforms. For example, high-fidelity gates can be implemented also via pulse sequences \cite{Casanova2015,Casanova2016, Casanova2017, Tratzmiller2021,Bartling2024} that can be tuned via experimental feedback calculated through our protocol. Furthermore, other defects in diamond (such as Group-IV centers) have recently gained significant interest. In such systems, nuclear-spin control~\cite{Stas2022,sukachev2017,Karapatzakis2024,Grimm2025} and single-shot readout of the nuclear spins~\cite{Klotz2025,Gundlapalli2025} have already been achieved. Due to the similar structure of the spin register in Group-IV defects and NV centers, an adaption of our findings will be straightforward.

\begin{acknowledgments}
    We acknowledge fruitful discussions with S. Montangero and S. G. Walliser.
	This work was supported by Germany’s Excellence Strategy – Cluster of Excellence Matter and Light for Quantum Computing (ML4Q2) EXC 2004/2 – 390534769, and by the European Union’s HORIZON Europe program via projects SPINUS (No. 101135699) and OpenSuperQPlus100 (No. 101113946) and by AIDAS-AI, Data Analytics and Scalable Simulation, which is a Joint Virtual Laboratory gathering the Forschungszentrum J\"ulich and the French Alternative Energies and Atomic Energy Commission. We also acknowledge support from the Marie Sklodowska-Curie Grant No. 101072637 (Project Quantum-Safe-Internet).
    We acknowledge the German Federal Ministry of Research, Technology and Space (BMFTR) for support via the projects SPINNING (No. 13N16210 and 13N16215), DE BRILL (No. 13N16207), Quanten4KMU (No. 03ZU1110BB), QuantumHiFi (No. 16KIS1593), QR.N, CoGeQ (13N16101) and Deutsche Forschungsgemeinschaft (DFG) via project No. 387073854, joint DFG-Japan Science and Technology Agency (JST) project ASPIRE (No. 554644981), Carl Zeiss Stiftung, and Ministry of Economic Affairs Baden-Württemberg via the project QC4BW.
\end{acknowledgments}

\section*{Authors contributions}
A.M. and M.M.M. conceived this research project. A.M. developed the formulation of the protocol and the simulations with the support of P.J.V. and M.M.M. A.M., P.J.V. and M.M.M. wrote the manuscript. 
All authors contributed to the discussion of the results and the manuscript.

\section*{Data availability statement}
The data and code for simulations and figures in this work is available on Zenodo.org (Ref.~\cite{marcominiComputationalResources}).

\section*{CORRESPONDENCE AND REQUESTS FOR
MATERIALS}
Should be addressed to A.M. at: 
\\
\indent amarcomini@vqcc.uvigo.es.

\section*{COMPETING INTERESTS}
The authors declare no competing interests.

\newpage
\bibliographystyle{apsrev4-2.bst}
\bibliography{MyPaper}

\clearpage
\newpage
\onecolumngrid
\appendix
\section{Implicit form of the system Hamiltonian}\label{sec: hamiltonians derivation}

From the considerations in Sec.~\ref{sec: System} and the results in Eqs.~\eqref{eq:system_Hamiltonian}-\eqref{eqn: spin approx}, the system is characterized by its full Hamiltonian 
\begin{equation}\label{eqn: H_cs}
 H =H_\text{d} + H_{\text{hf}} + H_\text{c},\\
\end{equation}
where the drift Hamiltonian $H_d$ is defined in Eq.~\eqref{eqn: drift hamiltonian} and the hyperfine and control Hamiltonians are
\begin{eqnarray}
 H_{\text{hf}}=\begin{pmatrix} 
 0&0&0&0\\
 0&0&0&0\\
 0&0&-\frac{A_{zz}}{2}&-\frac{A_{zx}-iA_{zy}}{2}\\
 0&0&-\frac{A_{zx}+iA_{zy}}{2}&\frac{A_{zz}}{2}
 \end{pmatrix}\,,
 \quad \quad
H_\text{c}=\begin{pmatrix}
    0&-\frac{\gamma_n}{2}B_1(t)&\frac{\gamma_e}{\sqrt{2}}B_1(t)&0\\
    -\frac{\gamma_n}{2}B_1(t)&0&0&\frac{\gamma_e}{\sqrt{2}}B_1(t)\\
    \frac{\gamma_e}{\sqrt{2}}B_1(t)&0&0&-\frac{\gamma_n}{2}B_1(t)\\
    0&\frac{\gamma_e}{\sqrt{2}}B_1(t)&-\frac{\gamma_n}{2}B_1(t)&0
\end{pmatrix}\,.
\end{eqnarray}
Upon transforming $H$ into the rotating frame with respect to the drift Hamiltonian, for the hyperfine interaction we have $\tilde{H}_{\text{hf}} = e^{iH_\text{d}t}H_{\text{hf}}e^{-iH_\text{d}t}$, which yields the result in Eq.~\eqref{eqn: tilde_H_hf}.
As for the control Hamiltonian, we have
\begin{eqnarray}\label{eqn: partial derivation of tilde_H_hf}
\tilde{H}_{\text{c}} = e^{iH_dt}H_{\text{c}}e^{-iH_dt}=\begin{pmatrix}
    0&-\frac{\gamma_n}{2}B_1(t)e^{-i\omega_n t}&\frac{\gamma_e}{\sqrt{2}}B_1(t)e^{-i\omega_e t}&0\\
    -\frac{\gamma_n}{2}B_1(t)e^{i\omega_n t}&0&0&\frac{\gamma_e}{\sqrt{2}}B_1(t)e^{-i\omega_e t}\\
    \frac{\gamma_e}{\sqrt{2}}B_1(t)e^{i\omega_e t}&0&0&-\frac{\gamma_n}{2}B_1(t)e^{-i\omega_n t}\\
    0&\frac{\gamma_e}{\sqrt{2}}B_1(t)e^{i\omega_e t}&-\frac{\gamma_n}{2}B_1(t)e^{i\omega_n t}&0
\end{pmatrix}\,.
\end{eqnarray}
We now set the control magnetic field as $B_1(t)=\frac{\sqrt{2}}{\gamma_e}\Omega_{\text{mw}}(t)\cos(\omega_{\text{mw}}\,t+\phi_{\text{mw}}) - \frac{2}{\gamma_n}\Omega_{\text{rf}}(t)\cos(\omega_{\text{rf}}\,t+\phi_{\text{rf}})$, i.e., a microwave part with slowly-varying amplitude $\Omega_{\text{mw}}(t)$ and carrier frequency $\omega_{\text{mw}}$, and a radio-frequency part with slowly-varying amplitude $\Omega_{\text{rf}}(t)$ and carrier frequency $\omega_{\text{rf}}$. Moreover, we omit the action of the microwave on the nuclear-spin transitions as well as the action of the radio-frequency on the electron-spin transition, which are far detuned, and we neglect the fast oscillating terms at frequencies $\omega_{\text{mw}}+\omega_e$ and $\omega_{\text{rf}}+\omega_n$, with $\omega_e=D-\gamma_e B_0$ and $\omega_n=\gamma_n B_0$. By plugging everything in Eq.~\eqref{eqn: partial derivation of tilde_H_hf} and performing the rotating-wave approximation, we obtain
\begin{eqnarray}
\tilde{H}_{c} = \begin{pmatrix}
    0&\frac{\Omega_{\text{rf}}(t)e^{i\phi_{\text{rf}}}}{2}e^{-i(\omega_n-\omega_{\text{rf}}) t}&\frac{\Omega_{\text{mw}}(t)e^{-i\phi_{\text{mw}}}}{2}e^{-i(\omega_e-\omega_{\text{mw}}) t}&0\\
    \frac{\Omega_{\text{rf}}(t)e^{-i\phi_{\text{rf}}}}{2}e^{i(\omega_n-\omega_{\text{rf}}) t}&0&0&\frac{\Omega_{\text{mw}}(t)e^{-i\phi_{\text{mw}}}}{2}e^{-i(\omega_e-\omega_{\text{mw}}) t}\\
    \frac{\Omega_{\text{mw}}(t)e^{i\phi_{\text{mw}}}}{2}e^{i(\omega_e-\omega_{\text{mw}}) t}&0&0&\frac{\Omega_{\text{rf}}(t)e^{i\phi_{\text{rf}}}}{2}e^{-i(\omega_n-\omega_{\text{rf}}) t}\\
    0&\frac{\Omega_{\text{mw}}(t)e^{i\phi_{\text{mw}}}}{2}e^{i(\omega_e-\omega_{\text{mw}}) t}&\frac{\Omega_{\text{rf}}(t)e^{-i\phi_{\text{rf}}}}{2}e^{i(\omega_n-\omega_{\text{rf}}) t}&0
\end{pmatrix}\,,
\end{eqnarray}
Note that this expression corresponds to the one in Eq.~\eqref{eqn: H control tilde final}, with the definitions of $\Omega_{\text{mw,c}}(t)$ and $\Omega_{\text{rf,c}}(t)$ in Eq.~\eqref{eqn: def mw control}.

\end{document}